\documentclass[aps, amsmath, amssymb, showkeys, showpacs, twocolumn] {revtex4-1}
\usepackage[colorlinks=false]{hyperref}
\usepackage{graphicx}
\usepackage{dcolumn}
\usepackage{bm}
\begin{document}
\title{A model violating the Boltzmann distribution}
\author{Yong-Jun Zhang}
\email{yong.j.zhang@gmail.com}
\affiliation{Science College, Liaoning Technical University, Fuxin, Liaoning 123000, China}

\begin{abstract}
We study a model of an ionic conductor having interstitial ions that jump from site to site. The conductor is subject to an external electric field. According to classical mechanics, the ion number density follows the Boltzmann distribution. But according to quantum mechanics, we show that the Boltzmann distribution is violated by a factor of 2. And the violation provides a mechanism to explain the Haven ratio observed in some experiments.
\end{abstract}
\maketitle

\section{introduction}
In statistical mechanics, the Boltzmann distribution \cite{B_Gibbs} is used indiscriminately in two kinds of situations. In the first situation, it deals with a single particle (or a system) in contact with a thermal reservoir of temperature $T$ and says that the probability to find the particle with energy $\varepsilon$ is $f(\varepsilon)\propto e^{-\frac{\varepsilon}{kT}}$. We call this distribution the Boltzmann distribution of the first kind, which always holds because it is directly related to the definition of \textit{temperature}.

In the second situation, the Boltzmann distribution deals with many particles in an external potential field $V(x)$.
And it says \cite{Kubo} that for the equilibrium state the particle number density is $n(x)\propto e^{-\frac{V(x)}{kT}}$. We call this distribution the Boltzmann distribution of the second kind, which is an extension of the Boltzmann distribution of the first kind.

Meanwhile, the particle number density ought to be determined by the principle of detailed balance \cite{Tolman}, which alway holds because it directly relates to the nature of \textit{equilibrium}.
According to the principle of detailed balance, the particle number density $n(x)$ should be determined by $n(x)P_{x\to x^{\prime}}=n(x^{\prime})P_{x^{\prime}\to x}$ where $P_{x\to x^{\prime}}$ and $P_{x^{\prime}\to x}$ are the transition probabilities which should be derived from the microscopic dynamics of the particles. The Boltzmann distribution of the second kind does not reflect the microscopic dynamics of the particles. So it can be in conflict with the principle of detailed balance in some models.

We will go through three models and compare them. The first two models have microscopic dynamics of classical mechanics. The third model has microscopic dynamics of quantum mechanics. The first two models will produce the Boltzmann distribution of the second kind. The third model will produce a result violating it. Note that some other works \cite{yjzhang, Bogdanov} also suggested possible violations of the Boltzmann distribution of the second kind.

\subsection{dilute gas in gravitational field}
We first examine a dilute gas in a gravitational field. According to the barometric formula \cite{Berberan-Santos}, the molecular number density follows the Boltzmann distribution of the second kind. In fact, all this is a result of the principle of detailed balance.

\begin{figure}[htbp]                                      
        \includegraphics[width=6cm]{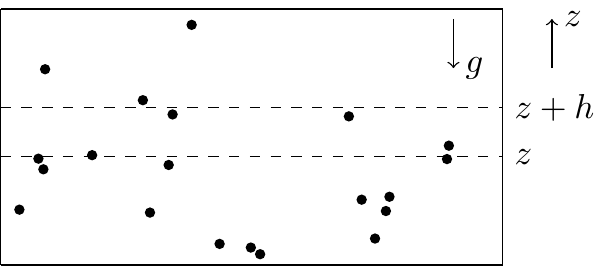}
\caption{
A dilute gas in a gravitational field. 
}\label{gas_random}                                           
\end{figure}

 Figure~\ref{gas_random} shows a dilute gas in a gravitational field. The dilute gas is in an equilibrium state.
Then the temperature is uniform \cite{Gibbs, cccc}, and the Maxwell-Boltzmann velocity distribution applies \cite{cccc,aaaa,bbbb}. Let us examine two cross sections at heights $z$ and $z\!+\!h$ and study only in the $z$-direction. This is a one-dimensional situation. The Maxwell-Boltzmann velocity distribution function becomes $f(v)\propto e^{-\frac{mv^2}{2kT}}$.

We focus on those processes involving no collisions. If a molecule at $z$ is able to reach $z\!+\!h$, it must have a velocity $v\geq \sqrt{2gh}$. So the corresponding transition probability is
\begin{equation} 
 P_{z\to z+h}\propto \int_{\sqrt{2gh}}^{\infty} v f(v)dv.
\end{equation}
If a molecule at $z\!+\!h$ is able to reach $z$ directly, its velocity just needs to point  downward. So the corresponding transition probability is
\begin{equation} 
  P_{z+h\to z}\propto \int_{-\infty}^{0} |v| f(v)dv=
 \int_{0}^{\infty} v f(v)dv.
\end{equation}
A calculation can show that
\begin{equation} 
 \frac{P_{z\to z+h}}{P_{z+h\to z}}= e^{-\frac{mgh}{kT}}.
\end{equation}
Putting it into the principle of detailed balance $n(z)P_{z\to z+h} = n(z+h) P_{z+h\to z}$, we get
\begin{equation} 
  \frac{n(z+h)}{n(z)} = e^{-\frac{mgh}{kT}}.
\end{equation}
This is the Boltzmann distribution of the second kind, and it remains the same when those processes involving collisions are taken into account \cite{zz}.
The same discussion can be applied to a dilute gas of charged particles in an electric field.

\subsection{ions in crystal in electric field}
The discussion about a dilute gas does not apply to interstitial ions in a rigid crystal. But if the microscopic dynamics of the ionic conduction is of classical mechanics, the principle of detailed balance still leads to the Boltzmann distribution of the second kind.

\begin{figure}[htbp]
        \includegraphics[width=3cm]{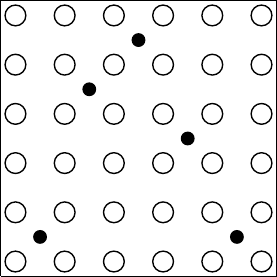}
\caption{
A simple solid ionic conductor. The open circles are the non-movable lattice ions. The filled circles are the movable interstitial ions that can jump from site to site.
\label{ions}}
\end{figure}

\begin{figure}[htbp]                                      
        \includegraphics[width=5cm]{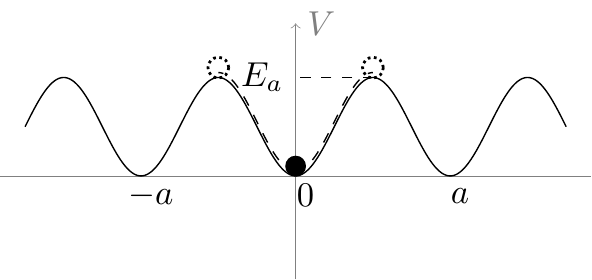}
\caption{
The classical mechanism for an ion to jump. The ion must have an energy high enough to surmount a peak of the surrounding potential.
}\label{peak}                                           
\end{figure}

Figure~\ref{ions} shows a solid ionic conductor.
Interstitial ions can jump from site to site. 
According to classical mechanics \cite{Kittel}, an ion can only jump when it can pass over a surrounding potential peak, as shown in Fig.~\ref{peak}.
So the ion must wait till it happens to have an energy high enough. 

\begin{figure}[htbp]                                      
        \includegraphics[width=5cm]{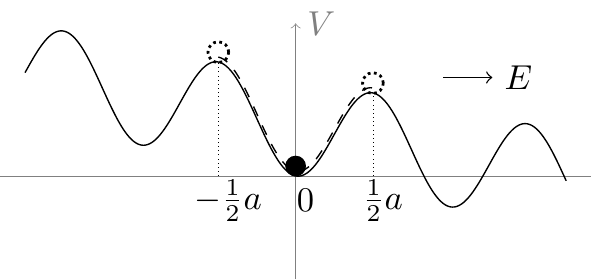}
\caption{
An electric field affecting the electric potential and affecting how an ion jumps. 
}\label{gap}                                           
\end{figure}

Before an electric field is applied, an ion has the same probability to jump to the right or the left. After the electric field is applied as shown in Fig.~\ref{gap}, the ion is more likely to jump to the right,
for which it just needs to have an energy $\varepsilon\geq E_a-\frac{1}{2}qaE$ where $E_a$ is the activation energy, $q$ is the ion's charge and $E$ is the electric field. Then the corresponding transition probability is
\begin{equation} 
 P_{0\to a}\propto \int_{E_a-\frac{1}{2}qaE}^{\infty} f(\varepsilon)d\varepsilon,
\end{equation}
where $f(\varepsilon)$ is the Boltzmann distribution of the first kind. For the ion to jump to the left, it needs to have an energy $\varepsilon\geq E_a+\frac{1}{2}qaE$. The corresponding transition probability is
\begin{equation} 
 P_{0\to -a}\propto \int_{E_a+\frac{1}{2}qaE}^{\infty} f(\varepsilon)d\varepsilon.
\end{equation}
A calculation can show that
\begin{equation}\label{secondstage} 
 \frac{P_{0\to -a}}{P_{0\to a}}= e^{-\frac{qaE}{kT}}.
\end{equation}
We can also introduce $P_{a\to 0}$. And obviously we have $P_{a\to 0}=P_{0\to -a}$. So we have $\frac{P_{a\to 0}}{P_{0\to a}}= e^{-\frac{qaE}{kT}}$.
Putting it into the principle of detailed balance $n_0P_{0\to a}=n_aP_{a\to 0}$, we get
\begin{equation} 
 \frac{n_0}{n_a}=e^{-\frac{qaE}{kT}},
\end{equation}
where $n_0$ and $n_a$ are the ion number densities at sites $0$ and $a$, respectively. This is the Boltzmann distribution of the second kind. But it will be violated if the microscopic dynamics of the ionic conduction is of quantum mechanics.

\section{Boltzmann distribution violated}
\begin{figure}[htbp]                                      
        \includegraphics[width=5cm]{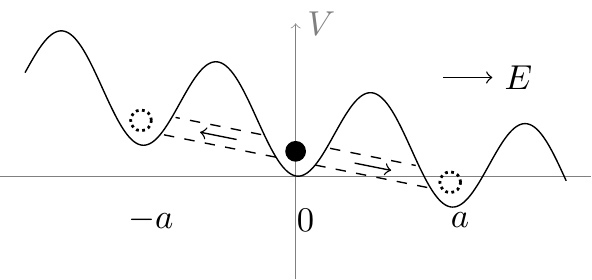}
\caption{
Quantum tunnelling. An ion can tunnel through the potential barrier no matter how low its energy is.
}\label{quantum}                                           
\end{figure}

The energy of an ion at a given site is quantized as Einstein proposed in his work \cite{Einstein_heat} about the specific heats of solids. 
So the ion is in a quantum state. Then according to quantum mechanics, the ion can take a \textit{jump} via quantum tunnelling \cite{Anderson, TLS}, as shown in Fig.~\ref{quantum}. 
Let us use $|0\rangle$ to denote the quantum state of the ion staying at the site 0 and in the ground state. Similarly, we introduce $|a\rangle$ and $|-a\rangle$. Then according to Fermi's golden rule, the probability per unit of time of the ion tunnelling from 0  to $a$ is
\begin{equation} \label{p0a}
 \Gamma_{0\to a} =\frac{2\pi}{\hbar} |\langle a|\hat{H}|0\rangle|^2\rho_f,
\end{equation}
where $\hat{H}$ (or $H^{\prime}$) is the Hamiltonian and $\rho_f$ is the density of states, which is determined by the conductor.
We denote the density of states of the conductor by $\rho_0$ for the most probable energy. Through the tunnelling process, the conductor will gain energy $qaE$, so its density of states will become $\rho_0 e^{\frac{aqE}{kT}}$ (see Appendix~\ref{aA}). So we have
\begin{equation}\label{D} 
  \Gamma_{0\to a} =\frac{2\pi}{\hbar}|\langle a|\hat{H}|0\rangle|^2 \rho_0 e^{\frac{qaE}{kT}}.
\end{equation}
Similarly, we get the probability per unit of time of the ion tunnelling from $0$ to $-a$,
\begin{equation} 
  \Gamma_{0\to -a} =\frac{2\pi}{\hbar}|\langle -a|\hat{H}|0\rangle|^2 \rho_0 e^{-\frac{qaE}{kT}},
\end{equation}
where the conductor loses energy $qaE$.

According to quantum mechanics, if an ion can tunnel from 0 to $a$, it can also tunnel from $a$ to 0. The two processes are time reversal to each other, and their Hamiltonian matrix element are complex conjugate to each other, $\langle 0|\hat{H}|a\rangle=\langle a|\hat{H}|0\rangle^*$. And, for a large conductor, we can use discrete translational symmetry to have $|\langle -a|\hat{H}|0\rangle|^2=|\langle 0|\hat{H}|a\rangle|^2$. So we have $|\langle -a|\hat{H}|0\rangle|^2=|\langle a|\hat{H}|0\rangle|^2$. Then we have
\begin{equation} 
 \frac{\Gamma_{0\to -a}}{\Gamma_{0\to a}}=e^{-2\frac{qaE}{kT}}.
\end{equation}
Then the ratio of the corresponding transition probabilities is
\begin{equation} 
 \frac{P_{0\to -a}}{P_{0\to a}}=e^{-2\frac{qaE}{kT}}.
\end{equation}
Let us further introduce $P_{a\to 0}$ and use $P_{a\to 0}=P_{0\to -a}$. Then we have $\frac{P_{a\to 0}}{P_{0\to a}}=e^{-2\frac{qaE}{kT}}$.
Putting it into the principle of detailed balance $n_0P_{0\to a}=n_aP_{a\to 0}$, we get
\begin{equation}\label{recover} 
 \frac{n_0}{n_a}=e^{-2\frac{qaE}{kT}}.
\end{equation}

If we take $x$-direction to be the opposite of the direction of the electric field and take the continuum limit, we have
\begin{equation}\label{extra2} 
 n(x)=n(0)e^{-2\frac{qEx}{kT}}.
\end{equation}
This distribution is different from the Boltzmann distribution of the second kind by a factor of 2.

We know that the Boltzmann distribution of the second kind leads to the Nernst-Einstein relation \cite{Einstein_relation, Kubo} $D=kT\mu$, where $D$ is the diffusion coefficient and $\mu$ is the ion mobility. Now that the Boltzmann distribution of the second kind is violated by a factor of 2, the Nernst-Einstein relation \cite{Kubo} would become
\begin{equation}\label{oneovertwo} 
 D=\frac{1}{2}kT\mu.
\end{equation}
In fact, it has been found in experiments that the Nernst-Einstein relation for ionic conduction is not exact, for which a parameter called Haven ratio has been introduced. The factor $\frac{1}{2}$ in relation (\ref{oneovertwo}) means a Haven ratio 0.5, which is consistent with most experiments \cite{Haven_ratio, Haven01, Haven03, Haven05, Haven06, Haven07, Haven08, Haven09, Haven10, Murch, Haven11, Haven17, Haven21, Haven22}. Though quantum tunnelling can happen beyond of the ground state, the result is the same (see Appendix~\ref{aB}).

\begin{acknowledgments}
The author is very grateful to Bin Zhang and Roberto Trasarti-Battistoni for many discussions. 
\end{acknowledgments}

\appendix
\section{\label{aA}density of states of the conductor }
Since the conductor is a macroscopic object, its energy eigenvalues can be considered as continuous. The energy of the conductor $\epsilon$ fluctuates around the most probable energy $\epsilon_0$ within a narrow range. We denote the conductor's density of states by $\rho(\epsilon)$ and use $\rho_0$ to stand for $\rho(\epsilon_0)$. Then by using the Boltzmann distribution of the first kind, we know that the probability to find the conductor to have energy $\epsilon$ is proportional to
\begin{equation} 
 \rho(\epsilon) e^{-\frac{\epsilon}{kT}}.
\end{equation}

The most probable energy $\epsilon_0$ is determined by $\frac{d}{d\epsilon}\left[\rho(\epsilon) e^{-\frac{\epsilon}{kT}}\right]=0$, which leads to
\begin{equation} 
 \frac{d\ln \rho(\epsilon_0)}{d\epsilon}=\frac{1}{kT}.
\end{equation}
So for a small change of energy $\Delta \epsilon$, we have
\begin{equation}\label{epsilon} 
 \rho(\epsilon_0 + \Delta \epsilon)= \rho(\epsilon_0)e^{\frac{\Delta \epsilon}{kT}} = \rho_0e^{\frac{\Delta \epsilon}{kT}}.
\end{equation}
Note that $\rho(\epsilon)$ is not necessarily the overall density of states of the entire conductor and can be just a relevant part of it.

\section{\label{aB}tunnelling in a general form}
We use $(0,n,i)$ to denote the quantum state of the ion at the site 0, where $n$ labels the energy eigenvalue and $i$ labels the eigenstate. Similarly, we introduce $(a, m, j)$ and $(-a, m,j)$. Note that all this is an approximation.

A quantum tunnelling may happen between any two eigenstates. For a given initial eigenstate $(0,n,i)$, we have
\begin{equation} \label{P0}
 \Gamma_{(0,n,i)\to (a,m,j)}= \frac{2\pi}{\hbar} |\langle a,m,j|\hat{H}|0,n,i\rangle|^2 \rho_0 e^{\frac{E_n-E_m+qaE}{kT}},
\end{equation}
and
\begin{equation}\label{Pma} 
 \Gamma_{(0,n,i)\to (-a,m,j)}=\frac{2\pi}{\hbar} |\langle -a,m,j|\hat{H}|0,n,i\rangle|^2 \rho_0 e^{\frac{E_n-E_m-qaE}{kT}},
\end{equation}
where $E_n$ and $E_m$ are the ion energy eigenvalues, which should be lower than the activation energy $E_a$.

For a specific situation where $m=n$ and $j=i$, we have
\begin{equation} 
  \langle 0,n,i|\hat{H}|a,n,i\rangle=\langle a,n,i|\hat{H}|0,n,i\rangle^*
\end{equation}
and
\begin{equation} 
  |\langle -a,n,i|\hat{H}|0,n,i\rangle|^2=|\langle 0,n,i|\hat{H}|a,n,i\rangle|^2.
\end{equation}
So we have
\begin{equation} 
 |\langle -a,n,i|\hat{H}|0,n,i\rangle|^2=|\langle a,n,i|\hat{H}|0,n,i\rangle|^2.
\end{equation}
To generate this specific relation, there must be a general relation
\begin{equation}\label{mam} 
 |\langle -a,m,j|\hat{H}|0,n,i\rangle|^2=|\langle a,m,j|\hat{H}|0,n,i\rangle|^2.
\end{equation}

Combining (\ref{mam}) with (\ref{P0}) and (\ref{Pma}), we get 
\begin{equation} 
 \frac{\Gamma_{(0,n,i)\to (-a,m,j)}}{\Gamma_{(0,n,i)\to (a,m,j)}}=e^{-2\frac{qaE}{kT}}.
\end{equation}
Then we can count and sum all the possible final states to get
\begin{equation} 
 \frac{\Gamma_{(0,n,i)\to (-a,\cdots) }}{\Gamma_{(0,n,i)\to (a,\cdots)}}= \frac{\sum\limits_{m,j}\Gamma_{(0,n,i)\to (-a,m,j) }}{\sum\limits_{m,j}\Gamma_{(0,n,i)\to (a,m,j)}}=e^{-2\frac{qaE}{kT}}.
\end{equation}
For the initial state of the ion at site 0, we introduce $p_{n,i}$ to denote the probability to find it in an eigenstate $(0,n,i)$. Then we have
\begin{equation} 
 \frac{P_{0\to -a }}{P_{0\to a}}=\frac{\sum\limits_{n,i} p_{n,i} \Gamma_{{(0,n,i)}\to (-a,\cdots) }}{\sum\limits_{n,i} p_{n,i} \Gamma_{(0,n,i)\to (a,\cdots)}}=e^{-2\frac{qaE}{kT}},
\end{equation}
which is independent of the probability distribution $p_{n,i}$.

\end{document}